\documentclass[12pt]{article}

\usepackage{amsmath}

\newcommand{\N}{N\raise.7ex\hbox{\underline{$\circ $}}$\;$}
\begin{document}

\title{Homogeneous space-times as models \\
 for isolated extended objects
\thanks{Report to Proceedings of 5th International Conference
Bolyai-Gauss-Lobachevsky: Non-Euclidean Geometry In Modern Physics
(BGL-5), 10-13 Oct 2006, Minsk, Belarus. }\\ }
\author{A. N. Tarakanov
\thanks{E-mail: tarak-ph@mail.ru }\\
{\small Minsk State High Radiotechnical College}\\
{\small Independence Avenue 62, 220005, Minsk, Belarus} }
\date{}
\maketitle
\begin{abstract}

An extended object is considered on the Minkowski background in
the form of a space-time bag, which is bounded by a certain
surface confining an internal substance. An internal metric is
built starting from the symmetry principles rather than from the
field equations. Assuming such a surface to be Lorentz invariant
we find that the internal space is proved to be the de Sitter
space. Conformal inversion of the internal metric relative to the
bag surface determines an external space (conformally conjugated
de Sitter space) whose metric may simulate a field of the object.
Although the extended object built in a such a way is noncompact,
its cross section by the hyperplane $r^0=0$, where $r^0$ is the
temporal coordinate, is compact (a ball) and the associated metric
can model a spherically symmetric extended massless charge in a
certain approximation.
\end{abstract}

\section{ Introduction}

\parindent=24pt\ \; \; \; Some times ago there were published papers
[1], proposing a model for hadrons, so called the Nijmegen model,
in which quarks and gluons were moving inside a closed space
domain, described by the anti-de Sitter (AdS) universe. This
theory may be considered as a variant of the bag model [2], which
is appeared to be fruitful for the spectral analysis of hadrons as
well as for a calculation of magnetic moments, decay widths,
electromagnetic mass splitting, etc. Experiments show that at
least heavy particles have an extended structure. To describe such
structure one may imagine hadron as a bag in the form of a certain
universe with the metric simulating interactions between quarks
and gluons considered as point objects. An indirect indication to
a character of these interactions may be the hadron spectroscopy.
In particular, an observation of spectra of $\psi$ and $\Upsilon$
particles, representable as excited states of quark-antiquark
systems $\it c\bar{c}$ and $\it b\bar{b}$, respectively, in which
forces between the constituents are described  by the oscillator
interaction, has lead to the AdS character of the bag interior.
For all that quarks are considered as point objects. Since an
exact interquark dynamics is not yet known, one may suppose such
point objects to be moving along the geodesic lines defined by the
internal bag geometry. Thus, for the description of the extended
particles in terms of the bag model one should use concepts and
methods of General Relativity.

\parindent=24pt\ It should be noted that the idea to describe
elementary particles in terms of General Relativity has rather
long history, but, to our mind, so far up to now is not properly
realized. The first attempt, probably, was made by Einstein and
Rosen [3] who had supposed the Schwarzschild solution for
describing neutral (neutron, neutrino) particles and the
Reissner-Nordstr\"{o}m solutions for charged particles. As early
as 1939 Mariani [4] had turned to metrical geometry containing
such a parameter in a natural way departing from that the theory
of elementary particles required introducing a fundamental length,
which cannot be taken into account by Euclidean geometry. At the
same time Lees [5] has proposed an electron model in the framework
of a consistent consideration of gravity and electromagnetism, in
which limitations on the electron shape and size were imposed, and
external field of the static electron was described by the
Reissner-Nordstr\"{o}m metric. Later Dirac has considered a
similar model of the electron [6] in the form of a pulsing bubble
with a surface tension. This model turned out to be a conceptual
basis of the bag model [2].

\parindent=24pt\ The most of different metrics, used in General
Relativity, are known to be obtained as solutions of either free
or material Einstein equations or their generalizations. At the
present time there exists no preferable choice between the metrics
applicable for a description of the same situation. This is due to
several circumstances: i) due to a physical interpretation of the
field equations and, in particular, to a meaning of the
energy-momentum tensor; a definition of the metric from the field
equations is not entirely correct procedure [7] (Ch. 7), all the
more that there exists ii) a problem of physical interpretation of
coordinates entering the particular expression for the metric [8].
Specifically, it causes an abundance of metrics, satisfying the
same equation, which may be obtained from each other by a certain
coordinate transformation. For the static vacuum spherically
symmetric fields this fact constitutes a content of the Birkhoff
theorem. iii) There exists also a widely discussed problem of
singularities of the space-time. On the one hand, the
singularities believed to be the coordinate effects and should be
eliminated from the theory. On the other hand, for example, a
supposition was being made that an existence of singularities,
probably, was a general property of all spaces, which might be
accepted as reasonable models of the Universe [9]. At any rate,
the surfaces of the metric singularities (horizons) are known to
play a great role in the black hole theory. Lastly, iv) a
relatively little number of geometro-physical experiments does not
give a possibility to make an unambiguous choice too.

\parindent=24pt\ {In our opinion, an invention of the metric
corresponding to a particular situation is possible without an
attraction of the field equations. The required properties of
symmetry and topology of the space-time, generated by congruences
of test-particles world lines treated as geodesics, a presence of
the singularities, determining the metric and topological
properties, as well as a coordinatization, connected with a
measurement process, should be more important circumstances than
any other ones. Moreover, the problems considered cannot be
exhausted with only gravity: other fields may also interpreted
from the viewpoint of the metric space-time properties [10] in the
spirit of the Poincar\'{e} conventionalism [11].}

\parindent=24pt\ {Among various spaces used at the present time,
there exists a class of maximally symmetric spaces, which, on the
one hand, may serve as a background for the evolution of physical
events and, on the other hand, they may be treated as a perfect
case of metrics, symmetry breaking of which to spherical, axial,
etc., one leads to metrics describing realistic situations. The
present paper considers just the case of maximal symmetry at
classical level. However, we built a metric starting from the
principles of the space-time symmetry rather than from the field
equations, which will be used for the interpretation of the metric
obtained. Assuming that the presence of a surface being a
space-time characteristic of an extended object implies a change
of original space-time geometry we shall try to answer a question:
what the metric must be, if test particles, whose world lines are
believed to be geodesic lines, are forbidden to come out of a
certain space-time domain for some reason. For the first
approximation, it is naturally to choose a three-dimensional ball
for a model of an isolated extended object. However, assuming if
this domain might have properties of maximal symmetry in the
perfect case (without of either a rotation of the object or a
presence of other objects or external fields) we choose not a
ball, but a region bounded by invariant surfaces as, for example,
a one-sheet hyperboloid and a light cone. Then a spherical
approximation may be obtained with the help of a suitable cross
section of it.}

\parindent=24pt\ {The paper is organized as follows. In Section 2
we depart from the Minkowski space splitted by invariant surfaces
into six regions. We require the line element to be invariant
along geodesic lines under coordinate transformations similar to
the velocity addition law in Special Relativity and obtain
expressions for the metrics of internal regions bounded by the
surfaces above. In Section 3 generators of coordinate
transformation groups, acting transitively in these regions, are
built, which form algebras so(2,3) or so(1,4). The metrics of the
spaces, bounded by a one-sheet hyperboloid and space-like infinity
of by a two-sheet hyperboloid and time-like infinity,
respectively, may be derived from the internal ones by the
conformal inversion relative to mentioned hyperboloids. The
spherical approximation is considered in Section 4. Finally,
Section 5 contents conclusive notes.}

\section{ Internal metric}

\parindent=24pt\ \; \; \; We start with the flat Minkowski space-time
${\bf E}_{1,3}^{\mathrm{R}}$ with the metric
$\eta_{\mu\nu}=diag(1,1,1,-1)$, however, we shall suppose that
test particle world lines for several reasons can neither come out
of some region $U\subset{\bf E}_{1,3}^{\mathrm{R}}$ nor penetrate
in it. In sufficiently large limits one may consider that
congruences of these world lines generate a Riemannian space, $\bf
V$, where they are geodesic lines [10]. The boundary $\partial U$
of the region $\it U$ divides $\bf V$ into internal, ${\bf
V}^{\mathrm{i}}$, and external, ${\bf V}^{\mathrm{e}}$, parts.
Thus, the region $\it U$ is represented by the Riemannian space
${\bf V}^{\mathrm{i}}$, which may be effectively considered as the
interior of an extended object with sharp boundary $\partial U$ in
the space-time ${\bf E}_{1,3}^{\mathrm{R}}$.

\parindent=24pt\ Starting from the principle of the general
coordinatization [12] we can use the Minkowski space
coordinatization as a coordinatization of internal, ${\bf
V}^{\mathrm{i}}$, and external, ${\bf V}^{\mathrm{e}}$, spaces
denoting their coordinates and metrics as $r^{\mu}$,
$h_{\mu\nu}(r^{\alpha})$ and $R^{\mu}$, $H_{\mu\nu}(R^{\alpha})$,
respectively. Then the line elements along geodesics are
\begin{equation}
  ds^{2}=h_{\mu\nu}(r^{\alpha})dr^{\mu}dr^{\nu}, \hspace{45mm}
\end{equation}
\begin{equation}
  dS^{2}=H_{\mu\nu}(R^{\alpha})dR^{\mu}dR^{\nu} \; , \qquad \alpha, \mu,
  \nu=0,1,2,3 \; .
\end{equation}
Obviously, the motions of both internal and external regions must
leave respective line element invariant, and the problem is to
seek out these motions depending upon a structure of the region U,
and vice versa.

\parindent=24pt\ Assuming the isolated extended object in question
to be described by Lorentz-invariant quantities let us consider
the boundary as Lorentz-invariant surface, or rather as a set of
Lorentz-invariant surfaces. Such surfaces are the one-sheet
hyperboloid, two-sheet hyperboloid and light cone, which are
defined as follows: \footnote{${\bf H}_{1,3}$ and ${\bf S}_{1,3}$
are equivalent to denotations ${\bf S}_{1}^{3}$ and ${\bf
H}_{0}^{3}={\bf H}^{3}$ in the monograph [13], respectively. See
also [14], [15].}
\begin{equation}
  {\bf H}_{1,3}\doteq \{{\it x^{\mu}}\in {\bf {E}}_{1,3}^{\mathrm{R}}: \eta_{\mu\nu}x^{\mu}x^{\nu}=-L^2\}=
  \textrm{SO(1,3)/SO(1,2)} \; ,
\nonumber
\end{equation}
\begin{equation}
  {\bf S}_{1,3}\doteq \{{\it x^{\mu}}\in {\bf E}_{1,3}^{\mathrm{R}}: \eta_{\mu\nu}x^{\mu}x^{\nu}=+L^2\}= \textrm{SO(1,3)/SO(3)} \;
  , \hspace{4mm}
\nonumber
\end{equation}
\begin{equation}
  {\bf C}_{1,3}\doteq \{x^{\mu} \in {\bf E}_{1,3}^{\mathrm{R}}:
  \eta_{\mu\nu} x^{\mu} x^{\nu}=0 \} =
  {\textrm{SO(1,3)}}/[{\textrm{SO(2)}} \times {\bf R}^1] \odot {\bf T}_2 \;
  ,\hspace{4mm}
\nonumber
\end{equation}
where ${\bf R}^1={\bf E}_{1}^{\mathrm{R}}$ is real straight line,
${\bf T}_2$ is translation group of two-dimensional plane ${\bf
R}^2 = {\bf E}_{2}^{\mathrm{R}}$, the sign $\odot$ denotes
semidirect product. Both ${\bf S}_{1,3}$ and ${\bf C}_{1,3}$ are
pairs of surfaces ${\bf S}_{1,3}^{\pm}$ and ${\bf C}_{1,3}^{\pm}$
with respect to $x^0 > 0$ and $x^0 < 0$. The surfaces defined in
such a way split the Minkowski space into six connected regions:

\parindent=24pt\ 1) interior of 1+3-dimensional pseudoball of hyperbolic type
\begin{equation}
  {{\bf D}_{1,3}^{\mathrm{H}}} \doteq \{x^{\mu} \equiv {r^{\mu}} \in {\bf E}_{1,3}^{\mathrm{R}}:
  -L^2 < \eta_{\mu\nu} r^{\mu} r^{\nu} \leq{0}\} \subset {\bf E}_{1,3}^{\mathrm{R}} \; ;
\end{equation}

\parindent=24pt\ 2-3) interiors of 1+3- dimensional pseudoball of spherical type
\begin{equation}
  {\bf D}_{1,3}^{+} \doteq \{x^{\mu} \equiv {r^{\mu}} \in {\bf E}_{1,3}^{\mathrm{R}}:
  0 < \eta_{\mu\nu} r^{\mu} r^{\nu} \leq {L^2}, \; r^{0} \geq {0}\} \subset {\bf E}_{1,3}^{\mathrm{R}} \; ,
\nonumber
\end{equation}
\begin{equation}
  {\bf D}_{1,3}^{-} \doteq \{x^{\mu} \equiv {r^{\mu}} \in {\bf E}_{1,3}^{\mathrm{R}}:
  0 < \eta_{\mu\nu} r^{\mu} r^{\nu} \leq{L^2}, \; r^{0} \leq {0}\} \subset {\bf E}_{1,3}^{\mathrm{R}} \; ,
\nonumber
\end{equation}
\begin{equation}
  {\bf D}_{1,3}^{\mathrm{S}} = {\bf D}_{1,3}^{+} \cup {\bf D}_{1,3}^{-} \; ;
\end{equation}

\parindent=24pt\ 4) exterior of 1+3- dimensional pseudoball of hyperbolic type
\begin{equation}
  {\bf \tilde{D}}_{1,3}^{\mathrm{H}} \doteq \{x^{\mu} \equiv {R^{\mu}} \in {\bf E}_{1,3}^{\mathrm{R}}:
  -\infty < \eta_{\mu\nu} r^{\mu} r^{\nu} \leq {-L^2}\} \subset {\bf E}_{1,3}^{\mathrm{R}} \; ;
\end{equation}

\parindent=24pt\ 5-6) exteriors of 1+3- dimensional pseudoball of spherical type
\begin{equation}
  {{\bf \tilde{D}}_{1,3}^{+}} \doteq \{x^{\mu} \equiv {R^{\mu}} \in {\bf E}_{1,3}^{\mathrm{R}}:
  {L^2} < \eta_{\mu\nu} r^{\mu} r^{\nu} \leq {\infty}, \; {R^0 > 0}\} \subset {\bf E}_{1,3}^{\mathrm{R}} \; ,
\nonumber
\end{equation}
\begin{equation}
  {{\bf \tilde{D}}_{1,3}^{-}} \doteq \{x^{\mu} \equiv {R^{\mu}} \in {\bf E}_{1,3}^{\mathrm{R}}:
  {L^2} < \eta_{\mu\nu} r^{\mu} r^{\nu} \leq {\infty}, \; {R^0 < 0}\} \subset {\bf E}_{1,3}^{\mathrm{R}} \; ,
\nonumber
\end{equation}
\begin{equation}
  {\bf \tilde{D}}_{1,3}^{\mathrm{S}} = {{\bf \tilde{D}}_{1,3}^{+}} \cup
  {{\bf \tilde{D}}_{1,3}^{-}} \; .
\end{equation}
($r^2 = \eta_{\mu\nu} r^{\mu} r^{\nu} = 0$ takes place in (3) and
(4) only at $r^0 = 0$).

\parindent=24pt\ As a point of departure for considering an
extended object we are interested mainly in the region ${\bf
D}_{1,3}^{\mathrm{H}}$ for its cross section by the hyperplane
$x^0 > 0$ represents a three-dimensional ball, which is the basis
for the representation of a spherically symmetric object. Then the
region ${\bf \tilde{D}}_{1,3}^{\mathrm{H}}$ may be treated as a
space, on which a force field of such an object is manifested. The
remaining regions are for the time being of purely geometric
interest, although all six ones may be considered by the same
method.

\parindent=24pt\ The metric in ${\bf V}^{\mathrm{i}}$, corresponding to ${\bf
D}_{1,3}^{\mathrm{H}}$ or ${\bf D}_{1,3}^{\mathrm{S}}$, may be
determined if we require the interval (1) to be invariant under a
transformation of coordinates $r^{\mu}$ satisfying condition (3)
or (4), respectively. Such a transformation may be written similar
to the velocity addition formula in Special Relativity [16]. Let
us consider a five-dimensional pseudo-Euclidean space ${\bf
E}_{2,3}^{\mathrm{R}}$ (or ${\bf E}_{1,4}^{\mathrm{R}}$) for the
case (3) (or (4)), covered by coordinates $\xi^{a}$,
$a=0,1,2,3,4$, with the line element
\begin{equation}
  ds^{2}=\eta_{ab} d\xi^{a} d\xi^{b} = \eta_{\mu\nu} d\xi^{\mu} d\xi^{\nu}
  + \eta_{55}({d\xi}^5)^2 \; ,
\end{equation}
where $\eta_{55} \equiv \eta = +1$ (or $-1$). Let us introduce now
in such space analogs of inertial frame systems (i.f.s.). Let
$\rho^{\mu}$ be the 1+3-dimensional analog of the relative
three-velocity between two i.f.s. Similarly to that the
coordinates of two i.f.s. in ${\bf E}_{1,3}^{\mathrm{R}}$ are
connected with each other by the Lorentz transformation, where the
three-velocity is a parameter, there takes place in our case an
analog of the Lorentz transformation for coordinates $\xi^{a}$:
$'\xi^{a} = \Lambda^{a}_{\; \cdot b} \xi^{b}$. A transformation
consistent with the condition (3) (or (4)) is the analog of
subliminal Lorentz transformation [16]:
\begin{equation}
  {\bf \Lambda} = {\bf 1} + \frac{\gamma}{L}{\hat{\bf \rho}}
  + \frac{\gamma - 1}{{L^2}{\beta^2}} {\hat {\bf \rho}}^2 \; , \qquad \gamma=(1-\beta^2)^{-1/2} \; ,
\end{equation}
where ${\hat {\bf \rho}} = \eta_{\mu\nu} \rho^{\mu} ({\bf
e}^{\nu5} - {\bf e}^{5\nu})$, ${\bf e}^{ab}$ are the elements of
complete matrix algebra satisfying to relations ${\bf e}^{ab} {\bf
e}^{cd} = \eta^{bc} {\bf e}^{ad}$, $({\bf e}^{ab})^{c}_{\cdot d} =
\eta^{ac} \delta^{b}_{\cdot d}$. The quantity
\begin{equation}
  \beta^2 = \frac{1}{2 L^2} {\mathrm{Sp}} ({\hat{\bf \rho}}^2) =
  -\frac{\eta}{L^2} \eta_{\mu\nu} \rho^{\mu} \rho^{\nu} =
  -\frac{\eta \rho^2}{L^2} \;
\end{equation}
runs the range $0\leq{\beta^2}<1$, whence it follows
$-L^2<\rho\leq{0}$ for the case (3) and $0\leq{\rho}<+ L^2$ for
the case (4). Thus, $\it L$ is here an analog of the velocity of
light. Coordinates $r^{\mu}$, $'r^{\mu}$ may be interpreted as
analogs of three-velocities $r^{\mu}=Ld\xi^{\mu}/d\xi^{5}$,
$'r^{\mu}=Ld'\xi^{\mu}/d'\xi^{5}$, connected by the transformation
of the velocity-addition law type
\begin{equation}
  'r^{\mu} = L\frac{\Lambda_{\cdot \nu}^{\mu} r^{\nu} + L\Lambda_{\cdot 5}^{\mu}}{\Lambda_{\cdot \nu}^{5} r^{\nu} + L\Lambda_{\cdot
  5}^{5}} = \frac{r^{\mu} + \rho^{\mu} + (\gamma-1)[1 + \frac{\eta_{\lambda\kappa} r^{\lambda} \rho^{\kappa}}{\rho^2}]}{\gamma[1 - \eta{\frac
  {\eta_{\lambda\kappa} r^{\lambda} \rho^{\kappa}}{L^2}]}} \; .
\end{equation}
Eq. (10) is a required transformation which ought to be
complemented by four-dimensional rotations of vectors $r^{\mu}$.
It it easy to see that $'r^{\mu}$ also satisfies the condition (3)
(or (4)) corresponding to $ds^2>0$. Thus, varying the parameters
$\rho^{\mu}$ and parameters responsible for four-dimensional
rotations of $r^{\mu}$ we obtain all points of the space ${\bf
D}_{1,3}^{\mathrm{H}}$ (or ${\bf D}_{1,3}^{\mathrm{S}}$). Defining
$d'r^{\mu}$ and $h_{\mu\nu}('r^{\alpha})$ from (10) we find from
the invariance condition of the interval (1) under the local
transformation (10) that the metrics $h_{\mu\nu}(r^{\alpha})$ have
a form
\begin{equation}
  h_{\mu\nu} =  \left (1+ \frac{\eta r^2}{L^2} \right )^{-1}\eta_{\mu\nu} - \left (1+\frac{\eta
  r^2}{L^2} \right )^{-2} \frac{\eta_{\mu\alpha}\eta_{\nu\beta}r^{\alpha}r^{\beta}}{L^2} \;
  ,
\end{equation}
\begin{equation}
  h^{\mu\nu} = \left (1+ \frac{\eta r^2}{L^2} \right )
  \left [\eta^{\mu\nu}+\frac{\eta r^{\mu}r^{\nu}}{L^2} \right ] \; ,
  \qquad h_{\mu\lambda}h^{\lambda\nu}=\delta_{\mu}^{\; \cdot \nu} \; .
\end{equation}
When $\eta = -1$ it is no more than the de Sitter (dS) metric
[17], which was initially obtained not from geometric
considerations but from the field equations with cosmological
term; $\eta = +1$ corresponds to anti-de Sitter (AdS) space.

\parindent=24pt\ It is easy to show that the coordinate transformation of the stereographic projection type [14]
\begin{equation}
  r^{\mu} = \left (1- \frac{\eta x^2}{4L^2} \right ) x^{\mu} \; , \qquad x^{2} =
  \eta_{\mu\nu}x^{\mu}x^{\nu} \; ,
\end{equation}
\begin{equation}
  x^{\mu} = -\frac{2 \eta L^2}{r^2} \; \left [1 + \sqrt{1+\frac{\eta r^2}{L^2}} \right ] \; r^{\mu} \; ,
  \end{equation}
reduces (1) to the standard expression for the line element of the
de Sitter space of constant curvature [13]
\begin{equation}
  ds^{2} = h_{\mu\nu} dr^{\mu} dr^{\nu} = \left(1 + \frac{\eta x^2}{4L^2} \right )^{-2}\eta_{\mu\nu} dx^{\mu} dx^{\nu} \; ,
\end{equation}
whence it follows for the Gaussian curvature $K = -\eta L^{2}$.
Hence, ${\bf D}_{1,3}^{\mathrm{H}}$ (${\bf D}_{1,3}^{\mathrm{S}}$)
is the space of constant negative (positive) and may be embedded
into the flat de Sitter space ${\bf E}_{2,3}^{\mathrm{R}}$ (${\bf
E}_{1,4}^{\mathrm{R}}$) with the metric (7), $\eta = +1$ ($\eta =
-1$). The metric (11) is the metric of the hypersurface
\begin{equation}
  \eta_{ab} \xi^{a} \xi^{b} = \eta_{\mu\nu} \xi^{\mu} \xi^{\nu} + \eta_{55} (\xi^{5})^{2} = \eta L^{2}
\end{equation}
in ${\bf E}_{2,3}^{\mathrm{R}}$ (or ${\bf E}_{1,4}^{\mathrm{R}}$)
written in terms of coordinates $r^{\mu}$, connected with
$\xi^{a}$ by the relations
\begin{equation}
  \xi^{\mu} = r^{\mu} (1 + \eta r^{2} / L^{2})^{-1/2} \; ,
  \qquad \xi^{5} = \pm L (1 + \eta r^{2} / L^{2})^{-1/2} \; ;
\end{equation}
\begin{equation}
  r^{\mu} = \xi^{\mu} (1 - \eta \xi^{2} / L^{2})^{-1/2} \; ,
  \qquad \xi^{2} = \eta_{\mu\nu} \xi^{\mu} \xi^{\nu} \; .
\end{equation}

\parindent=24pt\ In conclusion of this Section we note that just a
geometric approach was used by Einstein for a construction of the
metric of the closed static world [18]; as in is well known,
Friedmann [19] had generalized the Einstein's solution (as well as
the de Sitter's one [17]) to the non-static case.

\section{ The isometry group of internal space and external metric}

\parindent=24pt\ \; \; \; The transformation (10) together with four-dimensional
rotations forms an isometry group whose orbit is the whole space
${\bf V}^{\mathrm{i}}$ being thereby a homogeneous space. The
solution of the Killing equation
\begin{equation}
  h_{\nu\alpha} (\zeta_{A})_{; \; \beta}^{\nu} + h_{\nu\beta} (\zeta_{A})_{; \; \alpha}^{\nu} = 0 \; , \qquad A = \mu , \;
  [\lambda\kappa] \; ,
\end{equation}
gives four Killing vectors, $\zeta_{\mu}$, with components
\begin{equation}
  (\zeta_{\mu})^{\nu} = \delta_{\mu}^{\cdot \; \nu} + \eta \frac{\eta_{\mu\lambda} r^{\lambda} r^{\nu}}{L^2} \; ,
\end{equation}
responsible for transformations (10), and six Killing vectors,
$\zeta_{[\lambda\kappa]}$, with components
\begin{equation}
  (\zeta_{[\lambda\kappa]})^{\nu} = (\eta_{\lambda\alpha} \delta_{\kappa}^{\cdot \; \nu} - \eta_{\kappa\alpha} \delta_{\lambda}^{\cdot \; \nu})
  r^{\alpha} \; ,
\end{equation}
responsible for four-dimensional rotations of the vectors
$r^{\alpha}$. Thus, generators of the isometry group are
\begin{equation}
  {\hat{\bf J}}_{\mu} = -i L (\zeta_{\mu})^{\nu} \partial_{\nu} = -i L \left (\delta_{\mu}^{\cdot \; \nu} + \eta \frac{\eta_{\mu\lambda} r^{\lambda}
  r^{\nu}}{L^2} \right ) \partial_{\nu} \; ,
\end{equation}
\begin{equation}
  {\hat{\bf M}}_{[\lambda\kappa]} = -i L(\zeta_{[\lambda\kappa]})^{\nu} \partial_{\nu} =
  -i(r_{\lambda}\partial_{\kappa} - r_{\kappa}\partial_{\lambda})
  = \frac{1}{L} (r_{\lambda}{\hat{\bf J}}_{\kappa} - r_{\kappa}{\hat{\bf
  J}}_{\lambda}) \; .
\end{equation}

\parindent=24pt\ Assuming ${\hat{\bf J}}_{\mu} = {\hat{\bf M}}_{[\mu 5]} =
{\hat{\bf M}}_{[5 \mu]}$ we find the commutation relations between
${\hat{\bf M}}_{[ab]}$:
\begin{equation}
  [{\hat{\bf M}}_{[ab]},{\hat{\bf M}}_{[cd]}] = i (\eta_{ac} {\hat{\bf M}}_{[bd]} + \eta_{bd} {\hat{\bf M}}_{[ac]} -
  \eta_{ad} {\hat{\bf M}}_{[bc]} - \eta_{bc} {\hat{\bf M}}_{[ad]}) \; .
\end{equation}

\parindent=24pt\ Thus, as it should be expected an algebra of the
generators ${\hat{\bf J}}_{\mu}$, ${\hat{\bf
M}}_{[\lambda\kappa]}$ turns out to be isomorphic to the algebra
so(2,3) (for ${\bf D}_{1,3}^{\mathrm{H}}$) or so(1,4) (for ${\bf
D}_{1,3}^{\mathrm{S}}$), so that the isometry group O(2,3) acts
transitively on ${\bf D}_{1,3}^{\mathrm{H}}$ and O(1,4) acts
transitively on ${\bf D}_{1,3}^{\mathrm{S}}$. Moreover, SO(1,4) is
the group of isometry for ${\bf D}_{1,3}^{+}$. The group SO(1,3)
is a group of isotropy of the spaces in question, so that ${\bf
D}_{1,3}^{\mathrm{S}} = \textrm{SO(2,3)/SO(1,3)}$ and ${\bf
D}_{1,3}^{\mathrm{S}} = \textrm{SO(1,4)/SO(1,3)}$.

\parindent=24pt\ Commutation relations between ${\hat{\bf
J}}_{\mu}$'s show that the algebra of ${\hat{\bf M}}_{[ab]}$
becomes isomorphic to the Poincar\'{e} algebra iso(1,3) at the
limit $L\rightarrow\infty$. This is a reflection of the known fact
that the inhomogeneous pseudo-rotations group may be treated as
the limiting case of the homogeneous one in the space of one more
dimension [20]. The generators ${\tilde{\bf J}}_{\mu} = \hbar
L^{-1}{\hat{\bf J}}_{\mu}$ turn into $-i \hbar \partial_{\mu} =
{\hat{\bf P}}_{\mu}$ at the limit $L\rightarrow\infty$. However,
there exists also the limit $L \rightarrow 0$, when ${\tilde{\bf
J}}_{\mu} = \hbar L^{-1}{\hat{\bf J}}_{\mu} \rightarrow
\eta\eta_{\mu\lambda}r^{\lambda}r^{\nu} {\hat{\bf P}}_{\nu}$,
$[{\tilde{\bf J}}_{\mu}, {\tilde{\bf J}}_{\nu}] = 0$, and we come
again to the Poincar\'{e} algebra, but the metric (11) loses its
meaning because at this limit we have $h_{\mu\nu} \rightarrow 0$.
Formally, it could describe external regions ${\bf
{\tilde{D}}}_{1,3}^{\mathrm{H}}$ and ${\bf
{\tilde{D}}}_{1,3}^{\mathrm{S}}$ if one made a substitution
$r^{\mu} \rightarrow R^{\mu}$ and assumed $1 + \eta r^{2} / L^{2}
< 0$. However, the metric obtained in this way becomes
inconsistent with the condition above.

\parindent=24pt\ To derive an external metric $H_{\mu\nu}$ transforming
into $\eta_{\mu\nu}$ at $L \rightarrow 0$ it should be noted that
conditions (5), (6) may be derived from (3), (4) by the conformal
inversion of coordinates
\begin{equation}
  r^{\mu} = \frac{L^2}{R^2} R^{\mu} \; , \qquad R^{\mu} = \frac{L^2}{r^2} r^{\mu} \; ,
\end{equation}
with simultaneous transformation of the line elements
\begin{equation}
  ds = \frac{r^2}{L^2} dS = \frac{L^2}{R^2} dS
\end{equation}
specifying a one-to-one mapping from ${\bf V}^{\mathrm{i}}$ onto
${\bf V}^{\mathrm{e}}$. Going over to the coordinates $R^{\mu}$ in
(1) with the help of (25) and taking into account the relation
(26), we find that the metric of ${\bf V}^{\mathrm{e}}$ is
determined by
\begin{equation}
  H_{\mu\nu} = \left (1+ \frac{\eta L^2}{R^2} \right )^{-1}\eta_{\mu\nu} - \left (1+\frac{\eta
  L^2}{R^2} \right )^{-2}\frac{\eta {L^2} \eta_{\mu\alpha}\eta_{\nu\beta}R^{\alpha}R^{\beta}}{R^4} \; ,
\end{equation}
\begin{equation}
  H^{\mu\nu} = \left (1+\frac{\eta L^2 }{R^2} \right ) \left [\eta^{\mu\nu} +
  \frac{\eta {L^2} R^{\mu}R^{\nu}}{R^4} \right ] \; ,
  \qquad H_{\mu\lambda}H^{\lambda\nu}=\delta_{\mu}^{\; \cdot \nu} \; .
\end{equation}

\parindent=24pt\ Let us call a space-time with the metric (27) {\it the
conformally conjugated de Sitter space} (CCdS) or {\it the
conformally conjugated anti-de Sitter space} (CCAdS).

\parindent=24pt\ Coordinate transformation
\begin{equation}
  R^{\mu} = \frac{1}{2} \left (1 - \frac{\eta L^2 }{X^2} \right )X^{\mu} \; ,
  \qquad X^{2}= \eta_{\mu\nu}X^{\mu}X^{\nu} \; ;
\end{equation}
\begin{equation}
  X^{\mu} = -\frac{\eta R^2}{L^2} \left [1 \pm \sqrt{1 + \frac{\eta L^2}{R^2}} \right ] R^{\mu} \; ,
  \qquad \begin{pmatrix}
    {1} \\
    {0} \
  \end{pmatrix}
  \leq -\frac{\eta X^2}{L^2} <
  \begin{pmatrix}
    {\infty} \\
    {0} \
  \end{pmatrix} \; ,
\end{equation}
reduces the interval (2) to conformally flat form
\begin{equation}
  dS^{2} = H_{\mu\nu}dR^{\mu}dR^{\nu} = \frac{1}{4}
  \left (1 - \frac{\eta L^2 }{X^2} \right )^{4}
  \left (1 + \frac{\eta L^2 }{X^2} \right )^{-2} \eta_{\mu\nu}dX^{\mu}dX^{\nu} \; .
\end{equation}

\parindent=24pt\ A classification of all conformally flat
metrics by the isometry group is known [21] and therefore one may
say at once that the isometry group of ${\bf V}^{\mathrm{e}}$
coincides with its isotropy group, i.e. with the total Lorentz
group, moreover, a group of isometry for ${\bf
\tilde{D}}_{1,3}^{+}$ is a connected component of unit, SO(1,3).
It is easy verified directly by solving the Killing equation (19)
in the metric (27) giving six Killing vectors with components
(21).

\parindent=24pt\ According to embedding theorems [22] ${\bf V}^{\mathrm{e}}$
may be embedded into six-dimensional pseudo-Euclidean space ${\bf
\tilde{E}}_{2,4}^{\mathrm{R}}$ with the line element
\begin{equation}
  dS^{2} = \eta_{\mu\nu}dz^{\mu}dz^{\nu} + \eta (dz^5)^{2} - \eta
  (dz^6)^{2} \; .
\end{equation}
One may consider the metric (27) as the metric on
(2+4)-dimensional cone ${\bf C}_{2,4}$
\begin{equation}
  \eta_{\mu\nu}z^{\mu}z^{\nu} + \eta (z^5)^{2} - \eta
  (z^6)^{2} = 0
\end{equation}
in this space. Embedding formulae are
\begin{equation}
  z^{\mu} = \frac{\eta R^2}{L^2} \frac{\left [1 \pm \sqrt{1 + \frac{\eta L^2}{R^2}} \right ] R^{\mu}}{1 + \frac{\eta L^2}{R^2} \pm \frac{\eta R^2}{L^2} \sqrt{1 + \frac{\eta L^2}{R^2}}}
  = Z(R^2) R^{\mu} \; ,
\nonumber
\end{equation}
\begin{equation}
  z^{5} = \frac{Z(R^2)}{L} \left (R^2 - \frac{L^2}{4} \right ) \;
  , \; \;
  z^{6} = \frac{Z(R^2)}{L} \left (R^2 + \frac{L^2}{4} \right ) \; .
\end{equation}

\section{ Spherical symmetry}

\parindent=24pt\ \; \; \; A cross section of the one-sheet hyperboloid
$r^2 = R^2 = -L^2$ by the hyperplane $r^0 = 0$ is a sphere of the
radius $\it L$, being a singularity surface for related metrics
obtained from (11) and (27). Such a spherical approximation may be
of physical interest for both general relativistic and elementary
particles problems. Obviously, it is meaningful only for regions
${\bf D}_{1,3}^{\mathrm{H}}$ and ${\bf
\tilde{D}}_{1,3}^{\mathrm{H}}$, while it is possible for ${\bf
D}_{1,3}^{\mathrm{S}}$ and ${\bf \tilde{D}}_{1,3}^{\mathrm{S}}$
only at $|r^0| = {\mathrm{const}} > {\it L}$.

\parindent=24pt\ Spherically symmetric metrics corresponding to
${\bf D}_{1,3}^{\mathrm{H}}$ and ${\bf
\tilde{D}}_{1,3}^{\mathrm{H}}$ may be derived in two ways. First,
one may use eqs. (8)-(10), where $\rho^0 = 0$ should be assumed.
Then the consistent solution will be of the form (11) with
additional condition $r^0 = 0$. Just such a metric was used in
[1]. The corresponding external metric will have the form (27)
with additional condition $R^0 = 0$. Second, there may be assumed
simultaneously $\rho^0 = 0$, $r^0 = 0$ in the isometric
transformation (10) with $\bf \Lambda$ given by (8). Then metrics
$\hat{h}_{\mu\nu}$, $\hat{H}_{\mu\nu}$, derived in this way are
\begin{equation}
  \hat{h}_{0 \mu} = \hat{h}_{\mu 0} = \eta_{0 \mu} \; , \hspace{33mm}
  \qquad \hat{h}_{ij} = h_{ij}|_{r^0 = 0} \; ;
\end{equation}
\begin{equation}
  \hat{H}_{0 \mu} = \hat{H}_{\mu 0} = \eta_{0 \mu} \; ,
  \qquad \hat{H}_{ij} = H_{ij}|_{R^0 = 0} \; ;
  \qquad i,j=1,2,3 \; .
\end{equation}

\parindent=24pt\ A comparison of (35) with (11) and
(36) with (27) shows that distinctions occur only in
(00)-components. Respective line elements may be obtained from
each other by the transformation of differentials
\begin{equation}
  d\hat{r}^{0} = \left (1 - \frac{{\bf r}^2}{L^2} \right )^{-1/2} dr^{0} \; ,
  \qquad \hat{r}^{i} = r^{i} \; ,
\end{equation}
and
\begin{equation}
  d\hat{R}^{0} = \left (1 - \frac{L^2}{{\bf R}^2} \right )^{-1/2} dR^{0} \; ,
  \qquad \hat{R}^{i} = R^{i} \; ,
\end{equation}
where ${\bf r}^2 = - \eta_{ij}r^{i}r^{j} = (r^1)^2 + (r^2)^2
+(r^3)^2$, ${\bf R}^2 = - \eta_{ij}R^{i}R^{j} = (R^1)^2 + (R^2)^2
+(R^3)^2$.

\parindent=24pt\ Mentioned cross section causes a contraction
of the isometry groups of ${\bf V}^{\mathrm{i}}$ and ${\bf
V}^{\mathrm{e}}$ to ${\bf T} \times \mathrm{O(1,3)}$ and ${\bf T}
\times \mathrm{O(3)}$, respectively, where $\bf T$ is the
one-parameter group of time translations. In the first case
isometry groups act on arbitrary hyperplanes ${r^0} =
\mathrm{const}$, while in the case (35), (36) they act only on the
hyperplane $\hat{r}^0 = 0$. Thus, there arises a problem of a
choice, connected with a choice of a reference frame, between the
metric (35), (36) and the metrics derived in the first way, which
are in terms of spherical coordinates
\begin{equation}
  ds^{2} = \left (1 - \frac{{\bf r}^2}{L^2} \right )^{-1} \left [(dr^0)^2 - \left (1 - \frac{{\bf r}^2}{L^2} \right )^{-1} (d{\bf r})^2 - {{\bf r}^2}(d\vartheta^2 + {\sin^2 \vartheta})
  d\varphi^2 \right ] \; ,
\end{equation}
\begin{equation}
    dS^{2} = \left (1 - \frac{L^2}{{\bf R}^2} \right )^{-1} \left [(dR^0)^2 - \left (1 - \frac{L^2}{{\bf R}^2} \right )^{-1} (d{\bf R})^2 - {{\bf R}^2}(d\vartheta^2 + {\sin^2 \vartheta})
  d\varphi^2 \right ] \; .
\end{equation}
It should be noted that the external metric (40) at large ${\bf
R}^2$ may approximately be written as
\begin{equation}
    dS^{2} \approx \left (1 + \frac{L^2}{{\bf R}^2} + ... \right )
    (dR^0)^2 - \frac{1 + {3L^2}/{{\bf R}^2} + ...}{1 + {L^2}/{{\bf R}^2}+ ...}
    [(d{\bf R})^2 - {{\bf R}^2}(d\vartheta^2 + {\sin^2
    \vartheta}) d\varphi^2 ] \; .
\end{equation}
Hence, it follows that (40) approximates at very large distances
the Reissner-Nordstr\"{o}m metric for the massless charge $e
\approx L$. Formally assuming this charge to be the electron
charge we obtain for $\it L$: $L = \sqrt{\kappa \alpha h c}
\approx 7,6 \cdot 10^{-36} \; m$. Such interpretation, as well as
a choice between the metrics above, requires an additional
foundation based on the analysis of both field equations and
equations of motion. Here we only point out a form of the
energy-momentum tensor in cases (27) and (40). For the case (27)
we have
\begin{equation}
    G_{\mu\nu} = R_{\mu\nu} - {\frac{1}{2}}{H_{\mu\nu}}R = -\kappa
    T_{\mu\nu} = {\frac{12}{R^2}}{\eta_{\mu\nu}} -
    {\frac{3}{R^2}} \left (4 + {\frac{5 \eta L^2}{R^2}} \right )H_{\mu\nu} \; .
\end{equation}

\parindent=24pt\ Denoting the metric derived from (27) in the first way
as ${\bar{H}}_{\mu\nu}$ and the Einstein and energy-momentum
tensors through ${\bar{G}}_{\mu\nu}$ and ${\bar{T}}_{\mu\nu}$,
respectively, we find
\begin{equation}
    {\bar{H}}_{\mu\nu} = H_{\mu\nu}|_{R^0 = 0} \; , \hspace{86mm}
\end{equation}
\begin{equation}
    {\bar{G}}_{\mu\nu} = {\bar{R}}_{\mu\nu} - {\frac{1}{2}}{{\bar{H}}_{\mu\nu}}{\bar{R}} =
    -\kappa {\bar{T}}_{\mu\nu} =
    \hspace{60mm}
\nonumber
\end{equation}
\begin{equation}
    = G_{\mu\nu}|_{R^0 = 0} - {\frac{2 \eta L^2}{R^4}} \left [\eta_{\mu\nu} - \eta_{\mu
    0} \delta_{\cdot \nu}^{0} - \frac{1 + {\eta L^2}/{\bf R}^2}{1 - {\eta L^2}/{\bf
    R}^2} \; R^{i} {R^{j} \eta_{i \mu} \eta_{j \nu}} \right ] \; .
\end{equation}

\parindent=24pt\ As one can see from (42) and (44), an
interpretation of the metrics (27) and (40) is embarrassing and it
is not clear at the present time what are their sources. However,
in the physically interesting case of AdS spaces ($\eta = +1$) we
have the energy-momentum tensors with positive-definite
(00)-components, and $\bar{T}_{00} = T_{00} |_{R^0 = 0}$.

\section{ Conclusion}

\parindent=24pt\ \; \; \; A question of using the de Sitter groups in
General Relativity and elementary particles theory was being put
for a long time because they are unique minimal groups which may
contract (in the In\"{o}nu-Wigner sense) to the Poincar\'{e}
group. Thus, the de Sitter spaces may serve as models of a
physical space transforming into the flat one when curvature tends
to zero. In our case the de Sitter space appears as a result of
the consideration of the one-body problem. Indeed, if there is any
object in flat space-time, its field leads to an effective
distortion of the space what may be expressed in terms of
geometrical concepts as it was made above, for example. Here,
internal, ${\bf D}_{1,3}^{\mathrm{H}}$, and external ${\bf
\tilde{D}}_{1,3}^{\mathrm{H}}$, spaces are complementary for each
other (conformally conjugated). ${\bf D}_{1,3}^{\mathrm{H}}$
realizes the AdS geometry being considered comparatively rarely
because it has closed time-like geodesics. Assuming ${\bf
D}_{1,3}^{\mathrm{H}}$ to be a cosmological space with large
curvature radius, $\it L$, it is difficult to understand what one
has to do with such geodesics [23]. However, should one look at
from outside then such a behavior become consistent with the idea
of geometric confinement. Here we should like to pay attention to
one of possible approaches to introducing a fundamental length
basing upon a metrical geometry (see, e.g., [2]). In our case the
parameter $\it L$ can play a role of such a fundamental length.

\parindent=24pt\ In our opinion, an attractive feature of the metrics
obtained, (11) and (27) (and/or their spherically symmetric
"nonrelativistic" versions (35)-(36), (39)-(40)) is that they have
a unique irremovable singularity at $r^2 = R^2 = - L^2$ , i.e. on
the "surface" of the object in question. Therefore, one may
suggest the metric (27) (or (40)) for rough simulating the field
generated by an extended object with hyperbolic (or spherical)
symmetry placed into the Minkowski space-time ${\bf
E}_{1,3}^{\mathrm{R}}$. It enables one to give a clear
physico-geometric interpretation to coordinates $R^{\mu}$. Namely,
they are pseudo-Euclidean coordinates relative to the center of
symmetry $R^{\mu} = 0$ of points in which a field of the extended
object is detected. Certainly, at this point a question arises
about realisticity of the metric.

\end{document}